\documentclass[preprint,aps]{revtex4}
\usepackage{amsmath}
\input boxedeps.tex
\SetRokickiEPSFSpecial  
\HideDisplacementBoxes
\newcommand{\Tr}{{\rm \, Tr \!}}    
\newcommand{\half}{{1 \over 2}}

\def\lan{\langle}
\def\ran{\rangle}
\def\bra#1{\lan#1|}
\def\ket#1{|#1\ran}

\def\[{\begin{equation}}
\def\]{\end{equation}}

\def\ds{\displaystyle}

\begin{document}
\mbox{SWAT/396}\hfil\\
\mbox{SMUHEP/04-06}\hfil\\
\mbox{Revised May 2005}\hfil\\
\title{Spontaneously broken quark helicity symmetry}

\author{Simon Dalley}
\affiliation{Department of Physics, University of Wales Swansea \\
Singleton Park, Swansea SA2 8PP, United Kingdom}

\author{Gary McCartor}
\affiliation{Department of Physics, SMU  \\
Dallas, TX 75275}

\begin{abstract}

We discuss the origin of chiral symmetry breaking in the light-cone 
representation of QCD.  
In particular, we show how quark helicity symmetry is spontaneously 
broken in $SU(N)$
gauge theory with massless quarks
if that theory has a condensate of fermion lightcone zero modes.
The symmetry breaking appears as induced interactions
in an effective 
lightcone Hamiltonian equation based on a trivial vacuum.
The induced interaction is crucial for generating a splitting between 
pseudoscalar  and vector
meson masses, which we illustrate with spectrum calculations in  some
1+1-dimensional reduced models of gauge theory.

\end{abstract}

\maketitle


\section{Introduction}
\label{intro}

The lightcone Hamiltonian approach to quantum field  theories 
\cite{review} is an
intuitively appealing and powerful tool in the study of gauge theories.
There has been steady progress in recent years to formulate and solve
such theories at the non-perturbative level. In
pursuit of this goal, however,  one runs up against an apparent paradox.
One of the appealing features of lightcone quantization is the simplified
vacuum structure --- often it is chosen to be the trivial bare
vacuum. But in many cases of physical interest,  there is supposed to be a
complicated vacuum structure. The success of the QCD sum rules~\cite{sh}, 
for example,  suggests that
chiral and other condensates play an important role in extracting
hadronic physics from QCD. In this paper, we investigate how chiral
condensates could enter in the lightcone formulation of QCD and, assuming
that they are there, show how this leads to the breaking 
of quark  helicity symmetry with massless quarks. 
We derive the explicit form of the quark helicity violating
interactions, induced into
a lightcone QCD hamiltonian acting effectively  in a trivial vacuum.
It is shown to be necessary for splitting of pseudoscalar from 
vector meson masses and illustrative non-perturbative calculations
are performed in dimensionally reduced models.

The general mechanism by which a trivial light-cone vacuum can become dressed 
with fermions is now well understood \cite{mc94,nm00,mc00,mrp}, 
although detailed representations have been given 
only in a few simple models. In the cases that are completely understood, all 
of the physics of the full theory, including the dressed vacuum, can be 
represented by induced interactions that act in the usual 
light-cone representation space based on a trivial vacuum.  
In that sense, the problem is very similar to the standard renormalization 
problem:  degrees of freedom (those that dress the vacuum) are removed from 
the theory; when those degrees of freedom are properly "integrated out" 
rather than just ignored, new effects appear that provide the same physics 
using only the remaining degrees of freedom.  Of course, the usual 
high energy divergences of QCD also appear in the light-cone 
representation and must 
be controlled by some sort of regulators.  Counterterms associated
with those regulation procedures may also add new interactions to the theory.  
Given the interactions, a number of approaches
to renormalization of their couplings 
have been investigated, 
such as perturbative 
similarity renormalization 
group \cite{wilson} 
and non-perturbative symmetry optimization \cite{bvds}.

The fundamental requirement of Lorentz invariance implies that, 
{\em in the continuum}, 
the vacuum state should be the same in any quantization 
scheme.  Therefore, the vacuum in the light-cone representation is the same 
vector as in the `equal-time' representation, but the different 
quantization procedures give that vector in different representations.  
Likewise, condensates and any other quantity that is representation 
independent, must be the same whichever quantization scheme is used.  
When the theory is quantized directly on  null-planes,
the retention of lightcone zero
modes can dress the trivial vacuum and will do so if the theory 
incorporates a structured vacuum that supports condensates. This is the 
approach we follow here.
Let us note, however,  that there exist some 
other approaches to the lightcone vacuum 
problem which we do not pursue here, such as 
the use of near lightcone quantization \cite{near} or point-splitting in 
lightcone time \cite{point}. The spontaneous violation of helicity has also
been investigated in trivial bare lightcone vacua
perturbed by  mass breaking terms \cite{mass}. 
For any method,
computing the details of the representation 
of the vacuum in light-cone quantization of QCD would be quite involved. 
In this paper we have a more modest goal. Imposing the required
symmetries on the form of a general chiral-symmetry-breaking vacuum, we  
derive new effective interactions, operating in a trivial vacuum, 
that violate quark helicity.
The conservation of quark helicity
differs from that of chiral charge only by the neglect of zero mode
vacuum structure and hence
is sensitive to the spontaneous (and anomalous) breaking of chiral symmetry. 
The violation of quark helicity is crucial for
obtaining the correct hadron spectrum, previous 
literature behind this idea having been
reviewed by Mustaki \cite{must}.

The organization of this paper is as follows:
In the next section we introduce the  tools we will need, emphasizing
the role of helicity symmetry and rewriting
the theory in a way which exposes the 
zero mode operators that can dress the vacuum.
In section \ref{sec:QCD} we derive the form of  induced interactions that
violate quark helicity in massless $SU(N)$ gauge theory, assuming
the existence of a chiral symmetry-breaking fermion condensate. 
To illustrate the effect  on spectrum 
calculations, in particular 
pseudoscalar--vector splitting, we then perform some
computations on 1+1-dimensionally reduced models.
We obtain exact solutions to a model containing only the
new induced interactions in section~\ref{sec:exact}.
Adding gauge interactions, 
we then perform numerical DLCQ calculations 
of the meson spectrum.
Conclusions are discussed 
in section \ref{sec:last}. In Appendix A we briefly review the Schwinger Model,
which is an example where the fermionic vacuum structure is
exactly known \cite{nm00}.

\section{Helicity and the vacuum}
\label{sec:review}
\subsection{Quark helicity symmetry}
\label{sub:chiral}

Unless otherwise stated, our metric is such that lightcone coordinates are 
$x^\pm = x^0 \pm x^3$, $x_\perp = (x^1,x^2)$,
where $x^+$ is the light-front time variable. 
We decompose fermion spinors  $\psi = \psi_{+}
+\psi_{-}$ into the left and right moving 
projections $\psi_{\pm} = \frac{1}{2} \gamma_{\mp}
\gamma_{\pm} \psi$. 
Each of these
projections can be further decomposed into different helicities
 $\psi_{\pm,\uparrow} = \frac{1}{2} (1+\gamma_5) \psi_{\pm}$,
 $\psi_{\pm,\downarrow} = \frac{1}{2} (1-\gamma_5) \psi_{\pm}$.
Similarly, we define the helicity eigenstates of the transversely
polarized gauge fields via
 $A_{\uparrow} = (A_1 -{\rm i}A_2)/\sqrt{2}$, $A_{\downarrow} = 
(A_1 +{\rm i}A_2)/\sqrt{2}$.

In QCD with massless quarks, the action is
\begin{equation}
S = \int d^4 x  \left[ -{1\over 4} \Tr \ F_{\mu\nu} F^{\mu\nu}
+ {\rm i}\bar{\psi} \gamma^{\mu} D_{\mu} \psi 
\right] \ ,
\label{full}
\end{equation}
$\mu, \nu \in \{ 0,1,2,3\} $,  
\begin{equation}
D_{\mu} =  \partial_{\mu} + {\rm i}gA_{\mu}
\end{equation}
This action is invariant under the chiral transformation
\begin{equation}
\psi \to {\rm exp}(i\gamma_5 \theta) \psi \ .
\label{Xtran}
\end{equation} 
We work in light-cone gauge $A_- = 0$ throughout, 
in which case 
$\psi_-$ satisfies a constraint equation 
\begin{equation}
{\rm i} \partial_{-} \psi_{-} = {\rm i}\gamma_{\perp}\cdot D_{\perp}
\gamma^0 \psi_{+} \ .
\label{constraint}
\end{equation}
Integrating, we obtain
\begin{equation}
\psi_{-} = \psi_{-}^{0}(x^+,x^\perp) + \int dx^- \gamma_{\perp}\cdot D_{\perp}
\gamma^0 \psi_{+} \ .
\label{constrain}
\end{equation}
Here, $\int$ is the antiderivative which just replaces ${\rm e}^{{\rm i} k x}$ 
with 
${ 1 \over {\rm i} k} {\rm e}^{{\rm i} k x}$ in the Fourier expansion of the 
field.  We see that $\psi_-$ is not quite a dependent field of $\psi_+$.
The constant of integration is the  $x^-$-zero mode $\psi_-^0(x^+,x^\perp)$, 
which is a field that remains independent of $\psi_+$. This zero mode
can dress the lightcone vacuum state, breaking chiral symmetry, 
as we shall see.

\begin{figure}
\centering
\BoxedEPSF{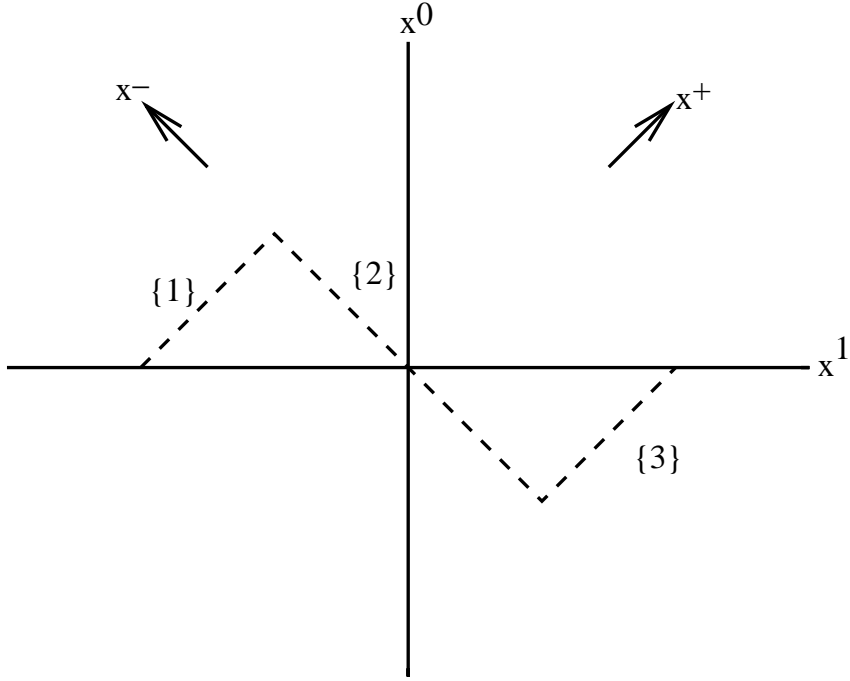 scaled 700}
\caption{A 
null surface for calculating charges.  Its extension to \{1\} and \{3\} is 
necessary to correctly incorporate fields that are independent of 
$x^-$. \cite{mcc}.}
\end{figure}

In order to cope with fields that may not diminish sufficiently at large
distances to discard, when necessary one can place the system in 
a box of length $2L$ in both $x^-$ and $x^+$ \cite{mcc}.  
If a charge is defined as the integral over $x^0 = 0$ of the 
zero component of a conserved current, one can calculate the same 
quantity by integrating over the 
surface illustrated
in Figure~1.  This includes not only the portion, labelled $\{2\}$ in 
figure~1, on which 
the fields are initialized in the light-cone representation, 
but also the legs $\{1\}$ and 
$\{3\}$.  In general the integrals over $\{1\}$ and 
$\{3\}$ do not go to zero as $L$ goes to infinity.  Integration constants 
such as $\psi_-^0$ cannot be initialized on surface 
$\{2\}$. Thus, for example,
on the $x^+ = 0$ initialization surface $\{2\}$ we have
\begin{equation}
\psi^{(a)}_{+,s}(0,x^-,x_\perp)   = {1\over\sqrt{\Omega_q}} 
\int_{0}^{\infty} dk^+
   b^{(a)}_s(k^+,x_\perp) {\rm e}^{-ik^+ x^-} +
   d^{(a)*}_{-s}(k^+,x_\perp,) {\rm e}^{ik^+ x^-} \label{fld1} . 
\end{equation}
Here
$(a)$ is the color index, $s$ 
the helicity index (such that if $s = \uparrow$, $-s = \downarrow$), 
and $\Omega_q$ a normalization
factor chosen so that the anti-commutation relations  are
\[
\{ b^{(a)}_{s_1}(k^+, x_{\perp}) ,b^{(b)*}_{s_2} (p^+, y_{\perp}) \} = 
 \delta(k^+ -p^+)\delta(x_{\perp} - y_{\perp}) \delta_{s_1 s_2} \delta_{ab}\ ,
\label{commf}
\]
with similar relations for anti-fermions $d$. 
The independent fermion $\psi_{-}^{0}$ is a function only of $x^+$ and may be 
expanded as 
\begin{equation}
\psi^{(0)(a)}_{-,s}(x^+, x_\perp)   = {1\over\sqrt{\Omega_q}} 
\int_{0}^{\infty} dk^-
   \beta^{(a)}_s(k^-,x_\perp) {\rm e}^{-ik^- x^+} +
   \delta^{(a)*}_{-s}(k^-,x_\perp,) {\rm e}^{ik^- x^+} \label{fld2} . 
\end{equation}
For gluons on $\{2\}$ we write
\begin{eqnarray}
    A^{(c)}_s(0,x^-,x_\perp) & = & 
{1\over\sqrt{\Omega_g}} \int_{0}^{\infty} dk^+ 
   \frac{1}{\sqrt{2 k^+}}\left(a^{(c)}_s(k^+,x_\perp) \, 
{\rm e}^{-i k^+ x^-} +
   {a^{(c)}_{-s}}^*(k^+,x_\perp,)  \,  {\rm e}^{i k^+ x^-}
 \right)  \ ,
\label{boson}
\end{eqnarray}
\[
[ a^{(b)}_{s_1}(k^+, x_{\perp}) ,a^{(c)*}_{s_2} (p^+, y_{\perp}) ] = 
 \delta(k^+ -p^+)\delta(x_{\perp} - y_{\perp}) \delta_{s_1 s_2} \delta_{ab}\ ,
\label{commg}
\]
Of course, for finite $L$ the
above integrals over $k^\pm$  become discrete sums for  integers $n$
of  $\pi n/ L$ (bosons) or $\pi (2n-1)/ 2L$ (fermions).

Of particular interest in the present case is the
chiral charge generator, defined by integrating the zero component
of the chiral current
\begin{equation}
j^{\mu}_5 = \bar{\psi} \gamma^\mu \gamma_5 \psi
\end{equation}
over $x^0 = 0$.  If the underlying chiral current is conserved, 
$\partial_{\mu} j^{\mu}_5 = 0$, then it is also equal to the integral of 
the normal component of the chiral current 
over $\{1\}$, $\{2\}$ and $\{3\}$.  If we neglect the 
contribution from surfaces at $x^- = \pm L$, one finds
\begin{eqnarray}
&& \int \bar{\psi} \gamma^+ \gamma_5 \psi \ d^2 x_{\perp}\  dx^- 
\label{lccharge} \nonumber \\
&& = \int dx_\perp dk^+ 
  \sum_{s} s [b^{(a)*}_s(k^+,x_\perp) b^{(a)}_s(k^+,x_\perp) 
   + d^{(a)*}_{s}(k^+,x_\perp) d^{(a)}_{s}(k^+,x_\perp)] 
\label{lchelicity}
\end{eqnarray}
which measures the total quark  and anti-quark 
helicity. Note that this is sometimes also called the 
light-front chiral charge 
\cite{must}.  This light-front chiral charge differs from the 
chiral charge by the 
integral of the chiral density over $\{1\}$ and $\{3\}$;  
$\psi_-^0$ will contribute to 
these last integrals.
If the underlying chiral current is conserved, 
and there are no non-trivial (vacuum) effects from zero modes on
the neglected surfaces, one expects quark helicity to be
conserved.
However, the non-conservation of quark helicity  is crucial to 
producing the experimentally observed
splitting in the meson spectrum between pseudoscalar and vector
mesons \cite{leut,wilson,perry}. 
To see this in more detail, one notes that the valence quark content 
of the pseudoscalar and helicity zero component of the vector is
$|\uparrow \downarrow\rangle \mp | \downarrow \uparrow \rangle$
respectively. They differ only by being symmetric and anti-symmetric
under quark helicity flip. Interactions that conserve quark helicity
cannot split them in energy. If the zero mode $\psi_-^0$ is neglected, 
the QCD action (\ref{full})
contains only quark helicity preserving interactions and cannot produce the 
required splitting. Therefore, we deduce that, in QCD, 
one cannot neglect the flow of chiral charge
across the surfaces at $x^- = \pm L$ and one must retain the zero modes that
live there. The violation of quark helicity is a direct consequence of
conserved chiral current and non-trivial structure of the vacuum
which violates spontaneously the chiral symmetry (\ref{Xtran}).

\subsection{Dressing the light-cone vacuum}
\label{sub:boson}

In order to investigate in more detail the considerations of the previous 
subsection, we now rewrite the fields in such a way as to expose 
the operators that can dress the vacuum. In  intermediate steps
in our analysis we will work with discretized $x_\perp$.
The fermion at given $x_\perp, a$, and  $s$ is then just a one 
dimensional fermi
field which can be bosonized in the form \cite{ls,naka}
\[
\psi^{(a)}_{+,s}(0,x^-,x^\perp) =  Z_+{\rm e}^{-\lambda_s^{(a)(-)}
(x^-,x_\perp)}
\sigma_{+,s}^{(a)}(x_\perp)
{\rm e}^{-\lambda_s^{(a)(+)}(x^-,x_\perp)} ,
\]
where $ Z_+$ is a wavefunction renormalization constant, 
\[
\lambda_s^{(a)(+)}(x^-,x_\perp) = -\int_{0}^{\infty} dk^+
 {1 \over k^+}C^{(a)}_{+,s}(k^+,x_\perp)({\rm e}^{-ik^+ x^-} - 
\theta(k - k^+)) ,   
\]
\[
\lambda_s^{(a)(-)}(x^-,x_\perp) = -{\lambda_s^{(a)(+)}}^* 
\]
\begin{eqnarray}
&&C^{(a)}_{+,s}(k^+,x_\perp)  =  
{\ds \int_{0}^{k^+} dq^+ d^{(a)}_{-s}\left(q^+ ,x_\perp\right) 
b^{(a)}_s\left(k^+ -q^+,x_\perp\right) + } \nonumber \\
&&  \int_{0}^{\infty} dq^+ b^{(a)*}_s \left(q^+,x_\perp\right) b^{(a)}_s
\left(k^+ + q^+,x_\perp\right) - \nonumber \\ 
&&  \int_{0}^{\infty} dq^+  d^{(a)*}_{-s} \left(q^+,k_\perp\right) d^{(a)}_{-s}
\left(k^+ + q^+ ,x_\perp\right) \label{c's} .
\end{eqnarray}
and $k$ is a Klaiber regulator \cite{klaiber}. Since the `fusion' fields $C_+$
are bosons,
 $\sigma_+$ is an $x^\pm$-independent
`spurion' inserted to  carry the fermionic quantum numbers of $\psi_+$.
It is important to appreciate that, in re-writing the fermi field
in this way, we have made no assumptions or approximations; the field 
written in terms of the $C$'s and $\sigma$'s is identical to (\ref{fld1}).
While we will not need most of the details of the fusion fields, the spurions
will play a key role.
We define a similar bosonization for the $\psi_-^0$ field in terms of
$\sigma_-(x_\perp)$ and $C_-(k^-,x_\perp)$. Since the fusion operator $C_-$
in this case is independent of $x^-$ ($k^+ = 0$), it will create
free  bosons of mass squared $-k_\perp^2$. These are tachyonic
and therefore cannot appear in the physical states of the theory,
although they may in general be needed to recover the 
full canonical structure. 
The operators we will 
retain from the $\psi_-^0$ fields are their `spurions' $\sigma_-$.  
For convenience, we note here the commutation relations of the spurions
that follow from those of the fermi field \cite{naka},
\begin{eqnarray}
\sigma_{\tau}^{*}  \sigma_{\tau}  = \sigma_{\tau}  \sigma_{\tau}^{*}
  & = & 1  \\
\{ \sigma_{\tau}^{*},  \sigma_{\rho}  \} = \{
\sigma_{\tau}  , \sigma_{\rho}  \} & = & 0 
\label{spur}
\end{eqnarray}
where $\tau$ and $\rho$ are differing labels indicating a 
Lorentz, colour, or transverse structure. Spurions 
 commute with the fusion operators.

At this point we can see that the presence of $\psi_{-}^{0}$, in particular
the spurion $\sigma_-$, allows for dressing of a trivial 
lightcone bare vacuum $|0\rangle$. 
Since the fusion operators $C_+(k^+)$ carry positive definite lightcone
momentum $k^+$, they cannot appear in the vacuum which should have momentum
zero (this is just the usual argument that leads to 
the trivial light-cone vacuum).
The spurions are kinematically allowed to dress the vacuum but 
considerations of symmetry restrict the possible ways that can occur. 
The fact that the vacuum must be a chargeless Lorentz scalar implies that,
in the sector where there are no gauge-field zero modes,
only bilinear combinations of the form $\sigma_{-,s}^* \sigma_{+,-s}$, or its
conjugate,  can occur. Note that these combinations, which are
the spurion part of $\bar{\psi} \psi$,  are not chirally invariant.
Other chargeless Lorentz invariant combinations of spurions can always
be reduced to combinations of the previous bilinear
forms using the rules (\ref{spur}).

The pure fermion content of  a general vacuum state 
must therefore be of the form
\begin{equation}
|\Omega_f\rangle = F[\sigma_{-,s}^{(a)*}(x_\perp) 
\sigma_{+,-s}^{(a)}(x_\perp),\sigma_{+,-s}^{(a)*}(x_\perp)
\sigma_{-,s}^{(a)}(x_\perp) ]
|0\rangle  \ .
\label{hoover}
\end{equation}
If the theory is to be charge conjugation invariant, the functional  
$F$ must be symmetric in its arguments. The spurion components
of the bilinears must also occur with the same colour and at the
same transverse position if the vacuum is to be invariant
under $x_\perp$-dependent gauge transformations
(we show later that 
$\sigma^{(a)}$ transforms in the fundamental representation when 
it acts on $|0\rangle$). Lastly, the functional $F$ should be
invariant under shifts of $x_\perp$ for translation invariance.

If $|\Omega_f\rangle$ is not equal to $|0\rangle$,
which we assume in this paper, 
then it follows that there is a $\bar{\psi} \psi$
condensate from this sector of the theory
\begin{equation}
\langle \bar{\psi} \psi(x_\perp) \rangle \sim 
\langle \Omega_f |  \sum_{s,a} \sigma_{-,s}^{(a)*}(x_\perp) 
\sigma_{+,-s}^{(a)}(x_\perp) + c.c.|\Omega_f\rangle  \ .
\label{cond}
\end{equation}
Translation invariance of the
condensate implies that the combination $\sigma_{-,s}^{(a)*}(x_\perp)
\sigma_{+,-s}^{(a)}(x_\perp)$ is independent of $x_\perp$ in the vacuum.
The contribution from $|\Omega_f\rangle$ may not be the total condensate, 
for sectors of the vacuum
containing gluon zero modes could also contribute, i.e. the total
vacuum is $|\Omega\rangle = |\Omega_f\rangle + |\Omega_g \rangle$, where
every vector in $|\Omega_g \rangle$ contains gluon fields. 
In fact, for the aim
of this paper in demonstrating quark helicity violation,
we will not need to consider the sector of the vacuum that 
includes gluon zero modes or 
the induced interactions that might result from that structure. That is not to 
say these sectors
are physically unimportant, but gluon condensates
and the effective interactions they generate is a separate
question to the one we address here. 
Not only do we not need the sector of the vacuum containing
gluon zero modes, but to find the form of the induced interaction
that violates quark helicity we shall not need to specify the exact form of 
$F$. There are examples of much simpler field theories where this vacuum 
functional can be obtained exactly. The most completely 
solved example is the Schwinger Model, for which 
we include a brief review in appendix A. For the case of the massive Schwinger 
model, there is an induced interaction whose form is exactly known \cite{mc99}.
That example serves to
demonstrate that the form (\ref{hoover}) can lead to 
induced interactions and, moreover, is sufficiently general to include the
case of anomalous chiral symmetry breaking and $\theta$-vacua.

\section{Induced Interactions in QCD}
\label{sec:QCD}

The lightcone QCD Hamiltonian $P^-$ derived from the action
(\ref{full}) satisfies an eigenvalue equation which is the 
relativistic Schrodinger equation for this problem: 
\begin{equation}
(P^+P^- - P_\perp^2) |\Psi\rangle = M^2 |\Psi\rangle \ .
\label{eig}
\end{equation}
$M$ is the invariant mass eigenvalue.
The eigenstates, $|\Psi\rangle$, can represent the vacuum, boundstates of
quarks and gluons (hadrons), or scattering states of hadrons. 
Since we focus on the new effects introduced by fermion lightcone
zero modes, we
do not need all the details of $P^-$. The interested
reader is referred to a review such as ref.\cite{review} 
for a fuller description
of the lightcone QCD hamiltonian.

Any interaction in $P^-$ that depends on $\psi_-$ can potentially give rise 
to an induced interaction through the existence of $\psi_{-}^{0}$.  
For the massless QCD case we consider in this paper,  in lightcone gauge
the only one is $-\bar{\psi} \gamma _{\perp} i D_\perp \psi $.  
The part of this interaction that depends on $\psi_{-}^{0}$ 
induces  extra, quark-helicity-violating, 
operators in the
QCD action (\ref{full}) and hence the lightcone hamiltonian. It is given by
\begin{eqnarray}
  I  & = &    I_1 + I_2 \nonumber \\
I_1 & = &        \int dx^-\;d^2x^\perp \sum_{a}
( i\partial_{\uparrow} \psi_{+,\downarrow}^{(a)*})  
\psi_{-,\downarrow}^{0 (a)} 
+ c.c.- [\downarrow \leftrightarrow \uparrow] \\
I_2 & = & g     \int dx^-\;d^2x^\perp 
\sum_{abc}\lambda_{a b}^c
 \psi_{+,\downarrow}^{(a)*}  \psi_{-,\downarrow}^{0 (b)} 
A_\uparrow^{(c)} + c.c.- [\downarrow \leftrightarrow \uparrow]
\label{eye}
\end{eqnarray}
where $\lambda_{a b}^c$ is the colour factor ($ab$ matrix element of the
$c^{\rm th}$ generator of $SU(N)$), 
\begin{equation}
\partial_{\uparrow} = (\partial_{1} -i \partial_{2})/\sqrt{2} \ ,  \   
\partial_{\downarrow} = \partial_{\uparrow}^*
\end{equation}
and $c.c.$ means complex conjugate.
We are going to study the
operators that result from the part containing 
only the spurion from the $\psi_{-}$ field. 
With that restriction, the first term of $I_2$, for example, 
is given by
\[
  g  Z_-  
\int dx^-\;d^2x^\perp  \sum_{abc} \lambda_{a b}^c
   {\psi}_{+,\downarrow}^{(a)*}
\sigma_{-,\downarrow}^{(b)}
{A}_\uparrow^{(c)}
\]
where  $Z_-$ is an (unknown)  renormalization constant that will depend
upon the regulator eventually used. 
From the identities (\ref{spur}), we can insert 1 in the form 
$\sigma_{+,\uparrow}^{(b)*}(x_\perp)\sigma_{+,\uparrow}^{(b)}(x_\perp)$, 
to rewrite it as
\[
 g  Z_-   \int dx^-\;d^2x^\perp  \sum_{abc} 
\lambda_{a b}^c {A}_\uparrow^{(c)}
{\psi}_{+,\downarrow}^{(a)*} \sigma_{-,\downarrow}^{(b)}
\sigma_{+,\uparrow}^{(b)*}
\sigma_{+,\uparrow}^{(b)}
.
\]
We now commute the spurions among themselves to get a combination on the 
far right, shown 
in square brackets below, that is in the same form as those appearing in the
pure-fermion part of the vacuum state (\ref{hoover}).
Proceeding in this way also for all the parts of $I_2$,  we  obtain:
\begin{eqnarray}
I_2 & = &  g Z_-  \int dx^-\;d^2x^\perp \sum_{abc}
\lambda_{a b}^c  \left\{
{A}_\uparrow^{(c)}
   {\psi}_{+,\downarrow}^{(a)*}
   \sigma_{+,\uparrow}^{(b)} 
   \Big[\sigma_{+,\uparrow}^{(b)*}\sigma_{-,\downarrow}^{(b)}\Big]
+
{A}_\uparrow^{(c)}{\sigma}_{+,\downarrow}^{(a)*} 
{\psi}_{+,\uparrow}^{(b)}\Big[\sigma_{-,\uparrow}^{(a)*}
\sigma_{+,\downarrow}^{(a)}\Big]
 \right. \nonumber  \\
&&  - 
\left. {A}_\downarrow^{(c)}{\sigma}_{+,\uparrow}^{(a)*}
{\psi}_{+,\downarrow}^{(b)} \Big[\sigma_{-,\downarrow}^{(a)*}
\sigma_{+,\uparrow}^{(a)}\Big]
 -
{A}_\downarrow^{(c)}
{\psi}_{+,\uparrow}^{(a)*}  \sigma_{+,\downarrow}^{(b)} 
\Big[\sigma_{+,\downarrow}^{(b)*}\sigma_{-,\uparrow}^{(b)}\Big] \right\} .
\label{f1}
\end{eqnarray}
Similarly, for $I_1$ we obtain
\begin{eqnarray}
I_1 & = &   Z_-  \int dx^-\;d^2x^\perp \sum_{a} 
\left\{    (i\partial_\uparrow {\psi}_{+,\downarrow}^{(a)*})
   \sigma_{+,\uparrow}^{(a)} 
   \Big[\sigma_{+,\uparrow}^{(a)*}\sigma_{-,\downarrow}^{(a)}\Big]
+
{\sigma}_{+,\downarrow}^{(a)*} 
(i\partial_\uparrow {\psi}_{+,\uparrow}^{(a)})
\Big[\sigma_{-,\uparrow}^{(a)*}
\sigma_{+,\downarrow}^{(a)}\Big]
 \right. \nonumber  \\
&&  - 
\left. 
{\sigma}_{+,\uparrow}^{(a)*}
(i\partial_\downarrow {\psi}_{+,\downarrow}^{(a)})
 \Big[\sigma_{-,\downarrow}^{(a)*}
\sigma_{+,\uparrow}^{(a)}\Big]
 -
(i\partial_\downarrow {\psi}_{+,\uparrow}^{(a)*})  \sigma_{+,\downarrow}^{(a)} 
\Big[\sigma_{+,\downarrow}^{(a)*}\sigma_{-,\uparrow}^{(a)}\Big] \right\} .
\label{f2}
\end{eqnarray}

We are now in a position to explain why the eigenvalue problem (\ref{eig}) 
can be reformulated with the trivial vacuum.  
The naive light-cone physical subspace consists of all polynomials in 
$\psi_+$ and $A_\perp$ applied to the 
light-cone bare vacuum $\ket{0}$.
We shall refer to this type of operator ${\mathcal O}_P$
as a physical operator and 
label the subspace of vectors ${\mathcal O}_P \ket{0}$ as $S_0$.  
The operators that will create physical states from the dressed vacuum are 
just these physical operators, $|\Psi \rangle = {\mathcal O}_P |\Omega 
\rangle$. $P^-= P^-_P + I + G_0$ will be the sum of a part, $P^-_P$, 
consisting entirely of physical 
operators, plus $I$, plus any other terms involving interactions induced by 
gluon modes in the vacuum.  Here we will set $G_0$ equal to zero and discuss 
the eigenvalue problem that results from including only the physical 
operators and $I$.
$P^+$ and $P^\perp$ are physical operators and are kinematical,
 so we may freely choose their value when we formulate the problem.  
We use
\begin{equation}
P^\perp = 0 \ \ \ , \ \ \ P^+=1 \ .
\end{equation}  
The eigenvalue equation (\ref{eig}) then takes the form  
\begin{equation}
(P^-_P + I){\mathcal O}_P|\Omega\rangle = 
M^2 {\mathcal O}_P|\Omega\rangle \ .
\label{eig2}
\end{equation}
From (\ref{f1})(\ref{f2}), 
we see that $I$ can be written in the general form
\begin{equation}
I \sim   I_{ P} [\sigma^*_{\pm} \sigma_{\mp}]
\end{equation}
where $I_{ P}$ is a physical operator and $[\sigma^*_{\pm} \sigma_{\mp}]$ 
is a bilinear
form occurring in $|\Omega_f\rangle$.  Suppose ${\mathcal P}$
projects onto $S_0$. Any physical operator commutes with ${\mathcal P}$.  
From the form  given in (\ref{hoover}), we see 
that any component of $\ket{\Omega}$, other than the bare vacuum $\ket{0}$,
must contain quanta from the $\psi^0_-$ field or gluon zero modes. 
Therefore, ${\mathcal P} \ket{\Omega} = c \ket{0}$ for some constant $c$.  
We also see that if an operator of the form 
$\sigma_{-,s}^{(a)*}(x_\perp) \sigma_{+,-s}^{(a)}(x_\perp)$ or its conjugate
acts on a component of 
$|\Omega_f\rangle$ in such a way as to remove 
all the $\sigma_-$ spurions, then 
it must also remove all the $\sigma_+$ spurions.
Therefore  ${\mathcal P} \sigma_{-,s}^{(a)*} (x_\perp)
\sigma_{+,-s}^{(a)}(x_\perp) 
\ket{\Omega} = \kappa \ket{0}$, where $\kappa$ is a real constant independent
of $s$, $a$, and $x_\perp$, as follows from Poincare and gauge invariance.

If we now operate on both sides of equation (\ref{eig2}) with ${\mathcal P}$ 
and divide through by $c$, we obtain
\begin{equation}
\left(P^-_P +   I_{P} {\kappa \over c} \right)
{\mathcal O}_P|0\rangle 
= M^2 {\mathcal O}_P|0\rangle \ .
\label{eig3}
\end{equation}
This equation has the same spectrum as (\ref{eig}) and the 
eigenvectors of (\ref{eig3}) are equal to the projections of the 
eigenvectors of (\ref{eig}) onto $S_0$.  Equation (\ref{eig3}) is an 
equation in 
the naive lightcone subspace.  
Of course, details of the 
dressed vacuum $| \Omega \rangle$ have been swept into the
unknown constants $\kappa$ and $c$.
In the following, we analyze the details of the action of $I_{P}$
on physical states, showing explicitly how it violates quark helicity.

For the projected equation (\ref{eig3}), the operators in $I_{P}$ are given
by (\ref{f1})(\ref{f2}) without the spurions in square brackets.
In order to demonstrate the action of ${I_P}$ on states in $S_0$, at this
point it is necessary to reintroduce the system in the box of Fig.1.
This will also facilitate later numerical work, where periodicity for bosons
and anti-periodicity for fermions over
$x^- \to x^- + 2L$   introduces a DLCQ \cite{dlcq} 
harmonic resolution cutoff $K = L/\pi$.  
Any state in $S_0$ can be written schematically (suppressing all
arguments) as
\begin{equation}
\sum_{i=1}^{\infty}\sum_{j=0}^{\infty} \sum_{k=0}^{\infty} f_{i,j,k} 
(b^{*})^i (d^*)^j (a^*)^k 
|0\rangle
\label{typical}
\end{equation}
To operate on such a state we can use the following properties of the spurions
 \cite{naka}
\begin{eqnarray}
&&\sigma_{+,s}^{(a)*} \ket{0}= 
{\rm e}^{{\rm i}  x^- \over 2K} b_{s}^{(a)*}(1/2K)
\ket{0} \\
&&\sigma_{+,s}^{(a)} \ket{0}= 
{\rm e}^{{\rm i} x^- \over 2K} d_{-s}^{(a)*}(1/2K)
\ket{0} \ 
\end{eqnarray}
which converts them to $\psi_+$ modes of the lowest allowed lightcone
momentum in DLCQ, $1/2K$. 
Note that this shows that $\sigma_{+}^{*}$ transforms in the
fundamental representation of the residual gauge group in lightcone gauge
when it acts on $\ket{0}$.
The following relations are also useful:
unless $s = s^\prime \,; \, a = b $ and $x_\perp = y_\perp$ $ 
\{\sigma_{+,s}^{(a)}(x_\perp) , b_{s^\prime}^{(b)*}(k^+,y_\perp) \; 
{\rm or} \; d_{-s^\prime}^{(b)*}(k^+,y_\perp)\} = 0$; otherwise
\begin{eqnarray}
&&\sigma_{+,s}^{(a)} b_{s}^{(a)*}((2n-1)/2K) = {\rm e}^{-{\rm i} x^- \over K}
b_{s}^{(a)*}((2n-3)/K) \sigma_{+,s}^{(a)} \\
&&\sigma_{+,s}^{(a)}d_{-s}^{(a)*}((2n-1)/2K) = {\rm e}^{{\rm i} x^- \over K}
d_{-s}^{(a)*}((2n+1)/2K) \sigma_{+,s}^{(a)} \\ 
&&\sigma_{+,s}^{(a)*} b_{s}^{(a)*}((2n-1)/2K) = {\rm e}^{{\rm i} x^- \over K}
b_{s}^{(a)*}((2n+1)/2K) \sigma_{+,s}^{(a)*} \\ 
&&\sigma_{+,s}^{(a)*} d_{-s}^{(a)*}((2n-1)/2K) = {\rm e}^{-{\rm i} x^- \over K}
d_{-s}^{(a)*}((2n-3)/2K) \sigma_{+,s}^{(a)*}\, ,
\end{eqnarray}
except that
\begin{eqnarray}
&&\sigma_{+,s}^{(a)} b_{s}^{(a)*}(1/2K) = {\rm e}^{-{\rm i} x^- \over K}
d_{-s}^{(a)}(1/2K) \sigma_{+,s}^{(a)} \\ 
&&\sigma_{+,s}^{(a)*} d_{-s}^{(a)*}(1/2K) =  {\rm e}^{-{\rm i} x^- \over K}
b_{s}^{(a)}(1/2K) \sigma_{+,s}^{(a)*} \, .
\end{eqnarray}

If we apply ${I_P}$ from $I_2$ (\ref{f1}) 
to a quark or antiquark in the state (\ref{typical}), 
we obtain
\begin{eqnarray}
    {I_P} \  b^{(a)*}_s(x+1/2K)
\ket{0}&  = & s g Z_-   \sqrt{\frac{\pi}{x}} \sum_{bc}\lambda^c_{ab} 
 \ b^{(b)*}_{-s}(1/2K) a^{(c)*}_s(x) \ket{0} \\
    {I_P}  \ d^{(b)*}_s(x+1/2K) \ket{0}
&= & -s g Z_-   \sqrt{\frac{\pi}{x}} \sum_{bc}\lambda^c_{ab}   
 d^{(a)*}_{-s}(1/2K) a^{(c)*}_s(x )\ket{0} 
\label{i1}
\end{eqnarray}
where $s =  +1 \implies \uparrow$, $s = -1 \implies
\downarrow $, and $x$ is a positive integer multiple of $1/K$.
The interaction puts the quark in the 
lowest DLCQ lightcone momentum state and flips its helicity. 
The same operator also produces pair annihilation:
\begin{eqnarray}
   {I_P} \; b_{s}^{(a)*}(x-1/2K) d_{s}^{(b)*}(1/2K) \ket{0} & = & 
   {I_P} \; b_{s}^{(a)*}(1/2K) d_{s}^{(b)*}(x-1/2K) \ket{0} \nonumber \\
& = & -s g Z_- \sqrt{\frac{\pi}{x}}  \sum_{c} 
\lambda^c_{ab} 
a_{s}^{(c)*}(x)\ket{0} \ .
\end{eqnarray}
Notice that the pair destruction only occurs when one 
quark is in the 
lowest DLCQ lightcone momentum state, 
so pair production always creates such a state.
If we apply ${I_P}$ from $I_1$ (\ref{f2}) 
to a quark or antiquark in the state (\ref{typical}), 
we find that it only acts on particles already in the 
lowest DLCQ lightcone momentum state. In this case, we
obtain
\begin{eqnarray}
    {I_P} \  b^{(a)*}_s(1/2K,x_\perp) \ket{0}
&  = & i s    Z_- \partial_s  b^{(a)*}_{-s}(1/2K,x_\perp)  \ket{0} \\
    {I_P}  \ d^{(a)*}_s(1/2K,x_\perp) \ket{0} &  = & 
 -i s Z_- \partial_s d^{(a)*}_{-s}(1/2K,x_\perp)  \ket{0} 
\label{i2}
\end{eqnarray}
The interaction flips the helicity of this quark
while changing  the `orbital' angular momentum of the state
to keep angular momentum projection $J_3$ conserved.

\begin{figure}
\centering
\BoxedEPSF{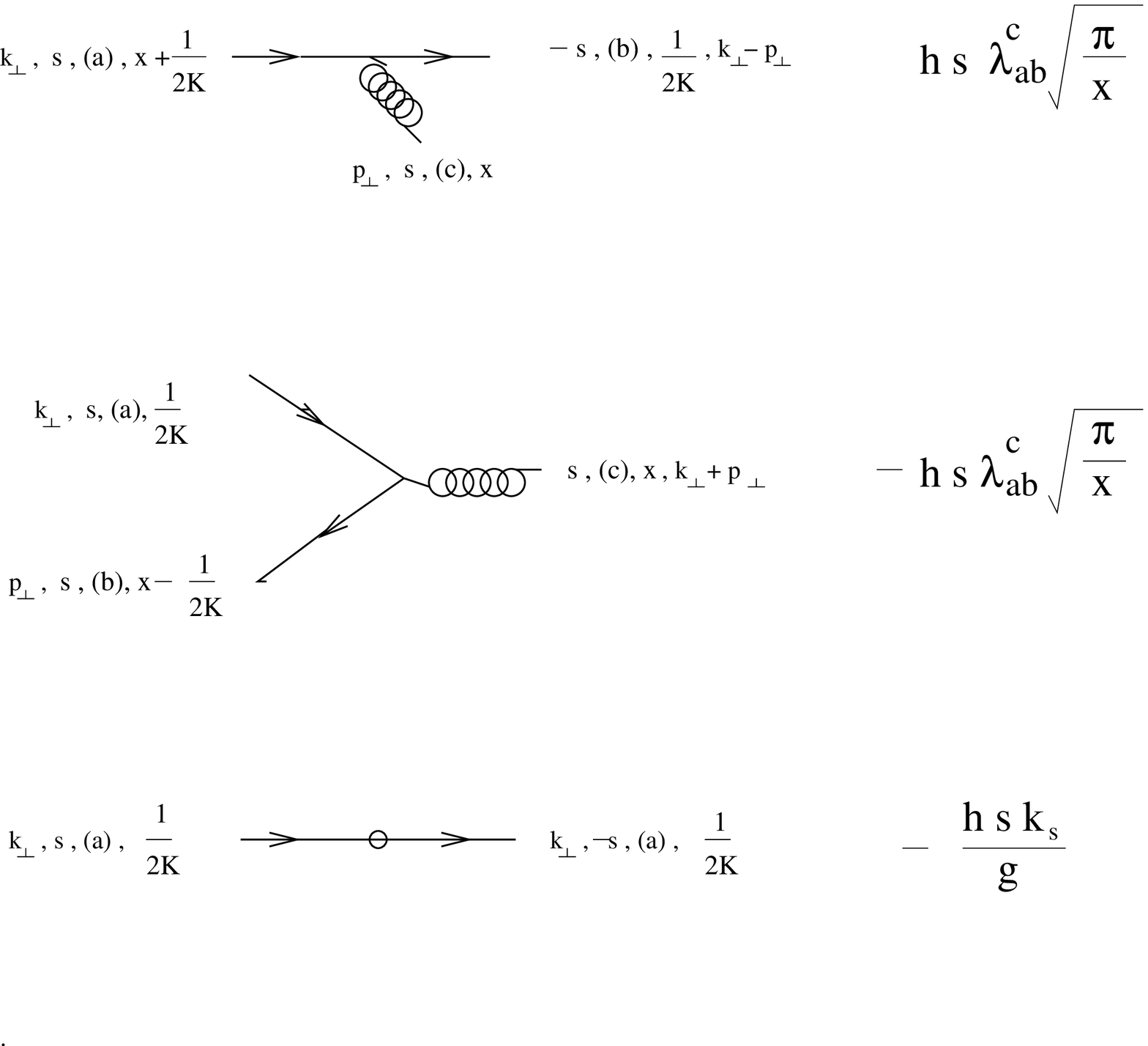 scaled 700}
\caption{The elementary vertices of lightcone perturbation
theory in momentum space, resulting from the induced operator $I$. 
They show the change of helicity $s$,
lightcone momentum $x$, colour $(a)$, and transverse momentum $k_\perp$,
where $k_s$ is the eigenvalue of $-i\partial_s$, $s \in \{\uparrow, 
\downarrow \}$. $1/2K$
is the smallest unit of fermion lightcone momentum in DLCQ
with cutoff $K$. $h$ is the combination of gauge coupling and vacuum
coefficients defined in eq.(\ref{defn}). 
The same vertices apply for
anti-quarks with arrows on quark lines reversed but with an extra overall
minus sign for the first and third amplitude (see eqs.(\ref{i1})(\ref{i2})).}
\end{figure}

Let us summarize the rather technical analysis of this section.
The net effect of the induced operator $I$ is in all cases to flip 
the helicity of a quark or anti-quark. Angular momentum is conserved
either by the emission of
 a gluon, leaving the quark in the lowest DLCQ momentum state,
 or a change in the orbital angular momentum of a quark already in the 
lowest DLCQ momentum state. 
Thus, quark helicity is no longer  conserved. 
The induced 
operator can also cause pair production of same-helicity quarks and 
antiquarks when one of them is in the 
lowest allowed
DLCQ momentum state. These processes are illustrated in Figure~2, where
we defined the effective coupling constant
\begin{equation}
h = {g Z_- \kappa \over c} \ . 
\label{defn}
\end{equation}
Evidently, all of the previous discussion goes through with any number
of spectator gluons and quarks in a hadron state.
The non-conservation of quark helicity is 
crucial to obtaining a splitting of masses between pseudoscalar
and vector mesons.
In the absence of
$I$ they would be degenerate. We mention that in a $U(N)$ gauge theory
the spin exchange in the valence state can proceed by 
annihilation through the color diagonal component of the gauge
field. In the case of QED, for example,  this contributes to the
hyperfine splitting of positronium. In a chiral $SU(N)$ gauge theory,
spontaneous (or anomalous) 
non-conservation of quark helicity is the only way to
achieve it.

\section{Solutions for Dimensionally Reduced Models}
\label{sec:exact} 

\subsection{Overview}

In this section,
we look at meson solutions for simple dimensionally reduced models in $1+1$ 
dimensions obtained by restricting to the $k_\perp = 0$ sector, meaning
that we (classically) discard all fields except the {\em transverse}
zero modes satisfying
\begin{equation}
\partial_{x_{\perp}} A_{\mu} = \partial_{x_{\perp}} \psi  = 0 \label{tzero} \ .
\end{equation}
These calculations, although not physically realistic, 
are straightforward enough to let us illustrate explicitly
the  effects on the meson spectrum alluded to above
and also allow us to investigate how things might scale with 
the DLCQ cutoff.  
In the present subsection, we find exact solutions for
the dimensionally reduced lightcone hamiltonian
containing only the induced operator $I$. In the next subsections
we analyze a slightly more realistic
large-$N$ gauged version of these dimensionally reduced models.  
In both cases, $I$ obviously consists of the part $I_2$ (\ref{f1}) only.

The  dimensionally reduced theories inherit various symmetries
from the $3+1$ dimensions.
They
preserve total quark and gluon helicity, identified with the angular momentum
projection $J_3$ for physical operators, a remnant of the 3+1 dimensional
rotation symmetry.
This automatically leads to doublets in the spectrum
consisting of opposite helicities.
There is also 
exact charge conjugation symmetry $C$ and
an exact kinematic parity
symmetry $P$ of the valence part of wavefunctions: 
$f^2 (x, 1-x) \leftrightarrow f^2 (1-x, x)$, where $x$ is the
quark lightcone momentum and $1-x$ the antiquark momentum . This kinematic
valence parity only equals parity of the  full wavefunction in
the free field limit, but is a convenient label. 
Thus, bound states can be labelled by $|J_3|^{PC}$. 
We will be particularly be interested in the quantum
numbers $0^{- \pm}$ and $1^{--}$, which together form 
the quantum numbers of a pseudoscalar, such as 
the $\pi$ or $\eta'$, and the three Lorentz components 
of a vector, such as the  $\rho$  or $\phi-\omega$ meson. Generically,  the 
$1^{--}$ doublet and the $0^{- -}$ are split because of 
dimensional reduction.
We do not address the issue here of how to make a full degenerate
3+1-dimensional Lorentz multiplet for the vector, as this really requires
transverse motion. Rather, we wish to study splitting of the $0^{-\pm}$ states,
since this occurs only if the quark helicity is not conserved. 
In $1+1$ dimensions, to have canonically normalised kinetic terms,
 we rescale fields $\psi \to \psi \sqrt{V_{\perp}}$, 
$A_{\perp} \to A_{\perp} \sqrt{V_{\perp}}$, where $V_{\perp} = \int dx^1 dx^2$
is the transverse volume factor.
The gauge coupling also becomes dimensionful through dimensional
reduction, $g^2 \to g^2 V_{\perp} N$, 
where we have also absorbed the colour factor
$N$ in the large $N$ limit.

\subsection{$I$ only}
\label{sub:I2}

In this subsection we shall consider the case where the entire lightcone
hamiltonian $P^-$ in the projected subspace $S_0$
is given by $I$ (\ref{eye}). 
In this case the colour label is rather redundant, so we
omit it.  
We can  expand the wave function of a helicity-zero `meson' as
\begin{eqnarray}
  \ket{\phi} &=& \sum_x f^{2}_{\uparrow\downarrow}(x) b_\uparrow^*(x)
d_\downarrow^*(1-x) \ket{0}\nonumber \\
 &+& \sum_{x+y \le 1} f^{3}_{\downarrow\uparrow\downarrow}(x,y) 
b_\downarrow^*(x)d_\downarrow^*(1-x-y)a_\uparrow^*(y) \ket{0} \nonumber \\
 &+& \sum_{x+y+z \le 1} f^{4}_{\uparrow\downarrow\uparrow\downarrow}(z,x,y) 
b_\uparrow^*(z)d_\downarrow^*(1-x-y-z)a_\downarrow^*(x) a_\uparrow^*(y)
\ket{0} \nonumber \\
&+& [\downarrow \leftrightarrow \uparrow] 
\label{expand} 
\end{eqnarray}
This expansion includes all the states which can couple to the `valence' $bd$
sector 
in this model, so it is a complete expansion of the wave function in the 
helicity zero sector.  The eigenvalue equation (\ref{eig3}), when projected
onto specific Fock states in the expansion (\ref{expand}), leads to the
following boundstate equations for the meson invariant 
mass $M$ 
\begin{eqnarray}
M^2 f^{2}_{\uparrow\downarrow}(x) & = & {h \over \sqrt{x}} 
f^{3}_{\downarrow\uparrow\downarrow}(0,x) 
\nonumber \\
& + &{h \over \sqrt{1-x}} f^{3}_{\uparrow\downarrow\uparrow}(x,1-x) 
\\
M^2 f^{3}_{\downarrow\uparrow\downarrow}(x,y) & = & {h \over \sqrt{y}} 
f^{2}_{\uparrow\downarrow}(x+y) 
\delta(x) 
\nonumber \\ 
& - &{h \over \sqrt{y}} f^{2}_{\uparrow\downarrow}(x) \delta(1-x-y) 
\nonumber \\ 
& - &{h \over \sqrt{x}} f^{4}_{\uparrow\downarrow\uparrow\downarrow}
(0,x,y) 
\nonumber \\ 
& + &{h \over \sqrt{1-x-y}} f^{4}_{\downarrow\uparrow\downarrow\uparrow}
(x,y,1-x-y) 
\end{eqnarray}
\begin{eqnarray}
M^2 f^{4}_{\uparrow\downarrow\uparrow\downarrow}(z,x,y) & = & -{h \over
 \sqrt{x}} f^{3}_{\downarrow\uparrow\downarrow}(z+x,y) \delta(z)
 \nonumber \\ 
& - &{h \over \sqrt{y}} 
f^{3}_{\uparrow\downarrow\uparrow}(z,x) \delta(1-y-z-x)
\end{eqnarray}
with the  same equations with 
$\uparrow \leftrightarrow \downarrow$, $h \leftrightarrow -h$.
0 is a shorthand for $1/2K$, and 1 for $1-1/2K$; 
if a denominator vanishes, then that term
is excluded from the equation; we define $\delta(x)=0$
if $x\neq 0$ otherwise 1.
This simple model is interesting only because
the equations can be solved algebraically.
They lead to  the following
effective equations in the valence sector.
\begin{eqnarray}
M^2 f^{2}_{\uparrow\downarrow}(0) & = & {h^2 \over M^2} 
\left( f^{2}_{\uparrow\downarrow}(0)-f^{2}_{\downarrow\uparrow}(1) \right)
\\
M^2 f^{2}_{\uparrow\downarrow}(x)|_{x \neq 0,1} & = & {h^2 \over M^2 x 
(1-x)} 
\left( 1+ {4 \over {M^4 \over h^2} - {1 \over x(1-x)}} \right)
f^{2}_{\uparrow\downarrow}(x)|_{x \neq 0,1}
\label{second} 
\end{eqnarray}
Solutions of the first of these equations for $f$ are delta functions at
0,1 and give the 
following results for the $0^{-+}$ state
\begin{equation}
f_{\uparrow\downarrow}(1)  =  f_{\uparrow\downarrow}(0) \ \ ; \ \ 
f_{\uparrow\downarrow}  =  -f_{\downarrow\uparrow}  \ \ ; \ \
M^2 =  \pm \sqrt{2} h
\end{equation} 
 and $0^{--}$ state
\begin{equation}
f_{\uparrow\downarrow}(1)  =  f_{\uparrow\downarrow}(0)  \ \ ; \ \ 
f_{\uparrow\downarrow}  =  f_{\downarrow\uparrow} \ \ ; \ \
M^2   =   0 \ . 
\end{equation}  
The masses are split even in the DLCQ continuum limit
$K \to \infty$. 

There are a number of 
artifacts that occur due to keeping only $I$ in this simple
model. The lowest lying states are tachyonic.
Opposite
parity states  ($f_{\uparrow\downarrow}(1) = \pm f_{\uparrow\downarrow}(0)$) 
are degenerate.
Also, there are solutions to eq.(\ref{second}) which are delta functions
at specific values $0<x<1/2$ with 
\[
M^4 = h^2 {1 \pm 2 \sqrt{x(1-x)} \over x(1-x)} \ .
\]
This spectrum  is continuous and, given that both signs are possible for $M^2$,
 unbounded below in the DLCQ
continuum limit $K \to \infty$. 
The unbound solutions at small
$x>1/2K$ are an artifact due to wee gluon emission. These artifacts will 
be avoided in the more realistic gauged model.

A similar analysis for the $1^{--}$ sector
produces $M^4 = h^2$ for delta function wavefunctions at $x = 0,1$. These
would partner the $0^{--}$ state and,
although they are not degenerate in the reduced model, would 
eventually form the vector Lorentz
multiplet in higher dimensions.

\subsection{Dimensionally reduced gauge theory}
\label{sub:dim}
Since the quark helicity violating term $I$ couples to the 
 quark of miminal DLCQ momentum only, one might
worry  that the effects shown in the last section 
disappear in the DLCQ continuum limit $L \to \infty$ of a non-trivial
theory that has wavefunctions with support on all momenta.
We demonstrate in the following with explicit  numerical
DLCQ calculations that this is not the case.
In this section, 
to be consistent with previous literature, we use the convention
$x^{\pm} =(x^0 \pm x^3)/\sqrt{2}$.
The light-cone hamiltonian we use starts from 
the large $N$ generalisation of eq.~(\ref{full}),
dimensionally reduced to 
$1+1$ dimensions,  with appropriately rescaled fields and coupling.
This results in a reduced  action that was first studied in ref.~\cite{anton}
\begin{eqnarray}
S & \to & \int dx^+ dx^- \ \left\{ -{1 \over 4} \Tr \ F_{\alpha \beta}
F^{\alpha\beta}+
{{\rm i} \over \sqrt{2}} (\bar{u}_{\uparrow} \gamma^{\alpha}_{(2)}
D_{\alpha} u_{\uparrow} + 
\bar{u}_{\downarrow} \gamma^{\alpha}_{(2)} D_{\alpha} u_{\downarrow}) 
 \right. \nonumber  \\
& & + \Tr\left[ -\half \bar{D}_{\alpha}
A_{r} \bar{D}^{\alpha} A^{r} -{{g}^2 \over 4N}
[A_{r},A_{s}][A^{r},A^{s}]
+ \half m_{0}^{2} A_{r}A^{r} \right]
\nonumber  \\
 & & \left. - {g\over \sqrt{2N}} (\bar{u}_{\uparrow} (A_1 + {\rm
i}\gamma^{5}_{(2)} A_2)
u_{\uparrow} - 
\bar{u}_{\downarrow}(A_1 - {\rm i} 
\gamma^{5}_{(2)} A_2 )u_{\downarrow}) \right\} \ .\label{red}
\end{eqnarray}
$\alpha$ and $\beta \in \{ +, -\}$, $r, s \in \{ 1,2 \}$,
$\gamma^{0}_{(2)} = \sigma^1$, $\gamma^{3}_{(2)} = {\rm i} \sigma^2$,
$\gamma^{5}_{(2)} = {\rm i} \sigma^1 \sigma^2$,
$\bar{D}_{\alpha} = \partial_{\alpha} + {\rm i}g[A_{\alpha},.]/\sqrt{N}$, 
$D_{\alpha} =
\partial_{\alpha} + {\rm i}gA_{\alpha}/\sqrt{N}$.
The two-component
spinors $u_{\uparrow}$ and $u_{\downarrow}$ are related to the original
$3+1$ dimensional $\psi$ field by
\begin{eqnarray}
2^{1/4} \psi \sqrt{\int dx^1 dx^2} = \left( \begin{array}{c}
 u_{+,\uparrow} \\  u_{-,\uparrow} \\  u_{-,\downarrow} \\  u_{+,\downarrow} 
 \end{array} \right)
\ , \
 u_{\uparrow} = \left( \begin{array}{c}
u_{-,\uparrow} \\ u_{+,\uparrow} 
\end{array} \right) \ , \ u_{\downarrow} = \left( \begin{array}{c}
u_{-,\downarrow} \\ u_{+,\downarrow} \end{array} \right) \ .
\label{spin}
\end{eqnarray}
Since the gluon mass is not protected by transverse
gauge symmetry transformations in a dimensionally reduced model, 
we must allow a gluon mass $m_0$ in general. 
In fact, this will regulate small-$x$ gluon divergences.
$u_{\uparrow},u_{\downarrow},A_1 ,A_2$ 
represent the transverse polarizations of the
$3+1$
dimensional quarks and gluons. In $1+1$ dimensions, where 
there is of course no spin, the fields appear as different `flavours' in 
fundamental and adjoint representations.

The $u_{\uparrow}$ and $u_{\downarrow}$ fields in (\ref{red})
have separate conserved $U(1)$ fermion numbers, but no axial
symmetries (with $\gamma^{5}_{(2)}$). 
This $U(1) \times U(1)$  transcribes to the left and right handed
$U(1)$  symmetries in $3+1$ dimensions of a single flavour of
massless quarks in QCD. Thus, the dimensionally reduced model has the important
property that it inherits the chiral symmetries of QCD with massless 
quarks. Note that, with a single flavour of quarks, the axial anomaly
may spoil one of the $U(1)$ symmetries, but in the large $N$ limit the anomaly
is suppressed as it involves fermion pair production. 
It is also necessary to work at large-$N$ to
have spontaneous symmetry breaking of continuous symmetry in $1+1$ dimensions.
It was shown 
in ref.\cite{alfaro}, by  completely different methods,
that this reduced theory
has spontaneous $U(1)$ symmetry breaking.  
The analogy with  $3+1$ dimensions at large $N$ 
and with one flavour, is that
one expects the axial $U(1)_{A}$ combination to be spontaneously 
broken. In the reduced model,
this corresponds to the $U(1) \times U(1)$ flavour symmetry being broken
down to its diagonal `total' fermion number subgroup. The broken $U(1)$
corresponds to the charge measuring quark helicity. 
Without the dimensionally reduced version of the induced
operator $I$, the hamiltonian contains no terms that flip quark
helicity. In that case, with a trivial vacuum, we find that the states 
$0^{-\pm}$ are degenerate. In the following, we add $I$ to the calculation
in the spirit of the last section, assuming there is a non-zero
condensate from the outset.

\subsection{Boundstate Equations}
\label{sub:bound}
In the
light-cone gauge, the fields
$A_+$ and non-zero modes of 
$u_{-}$ are non-propagating in light-front time $x^+ =
(x^0 + x^3)/\sqrt{2}$. We eliminate them using their
constraint equations of motion.
The expansion in creation and annihilation operators
for the dimensionally reduced fermion $u^{(a)}_{+,s}$
and gluon $A^{(c)}_s$ follows eqs.(\ref{fld1})(\ref{boson}) with
transverse momentum dependence dropped.
For a
helicity-zero meson in the projected subspace $S_0$, 
the state invariant under residual gauge transformations
is written 
\begin{eqnarray}
&& \sum_{a} 
\frac{1}{N}\int dx f^{2}_{\uparrow\downarrow} (x,1-x)  b^{(a) *}_{\uparrow}(x) 
d^{(a) *}_{\downarrow}(1-x) + 
\int dx f^{2}_{\downarrow\uparrow} (x,1-x) b^{(a) *}_{\downarrow}(x) 
d^{(a) *}_{\uparrow}(1-x)  \nonumber \\
&& + \sum_{abc} \frac{1}{N^2}
\int dx dy f^{3}_{\downarrow\uparrow\downarrow} (x,y,1-x-y) 
b^{(a) *}_{\downarrow}(x) \lambda_{ab}^{c} a^{(c) *}_{\uparrow}(y)
d^{(b) *}_{\downarrow}(1-x-y) +  \nonumber \\
&& \frac{1}{N^2} \int dx dy f^{3}_{\uparrow\downarrow\uparrow} (x,y,1-x-y) 
b^{(a) *}_{\uparrow}(x) \lambda_{ab}^{c} a^{(c) *}_{\downarrow}(y)
d^{(b) *}_{\uparrow}(1-x-y) + \cdots \ket{0} 
\label{meson}
\end{eqnarray}
where $\cdots$ indicates higher numbers of gluon creation operators
$a^{*}$. 
The wavefunction components are normalised as
\begin{eqnarray}
&& \int_{0}^{1} 
dx \ |f^{2}_{\uparrow\downarrow} (x,1-x)|^2 + |f^{2}_{\downarrow\uparrow}
 (x,1-x)|^2  \nonumber \\
&& + \int_{0}^{1} dx \int_{0}^{1-x} dy \ |f^{3}_{\downarrow\uparrow\downarrow} 
(x,y,1-x-y)|^2 + 
|f^{3}_{\uparrow\downarrow\uparrow} (x,y,1-x-y)|^2 + \cdots = 1
\end{eqnarray}

As before, boundstate equations for the wavefunctions $f$
are obtained by applying the
lightcone hamiltonian $P^-$ to a meson state, such as eq.(\ref{meson}), 
and then
projecting onto a given vector in the physical Fock space.
If we neglect contributions from gluon zero modes,
the resulting equations are  the same as those
given in ref.\cite{anton},
plus the helicity-violating
induced interaction $I$. 
Since we will solve these equations numerically in section \ref{sub:solve}, 
in a  truncation of the
Fock space to the sectors of $f^2$ and $f^3$ (one-gluon approximation),
we display only the equations for this truncation:
\begin{eqnarray}
{M^2 }f^{2}_{\uparrow\downarrow}(x,1-x) & = &  {m_{f}^2 \over x(1-x)}
f^{2}_{+-}(x,1-x)
\nonumber \\
& + &{{g}^2  \over \pi} \int_{0}^{1} dy
\left\{ {f^{2}_{\uparrow\downarrow}(x,1-x) - f^{2}_{\uparrow\downarrow}(y,1-y)
 \over (y-x)^2}\right\}
\nonumber \\
&+ & {h \over \sqrt{x}} f^{3}_{\downarrow\uparrow\downarrow}(0,x,1-x) 
\nonumber \\
& + &{h \over \sqrt{1-x}} f^{3}_{\uparrow\downarrow\uparrow}(x,1-x,0) 
\label{two}
\end{eqnarray}
\begin{eqnarray}
 M^2 f^{3}_{\downarrow\uparrow\downarrow}(x,y,1-x-y) & = &
{m_{b}^{2} \over y}  f^{3}_{\downarrow\uparrow\downarrow}(x,y,1-x-y)
\nonumber \\
&+ & {{g}^2 \over \pi}\int_{0}^{1-x} dz  {1+y-x-z
\over 2(1-x-y-z)^2\sqrt{y(1-x -z)}} \{
f^{3}_{\downarrow\uparrow\downarrow}(x,y,1-x-y) 
\nonumber \\
&&  -  f^{3}_{\downarrow\uparrow\downarrow}(x,1-x-z,z) \}  
\nonumber \\ &
+ & {{g}^2  \over \pi (1-x-y)}
\left(  \sqrt{1 + {1-x-y \over y}} -1 \right) 
 f^{3}_{\downarrow\uparrow\downarrow}(x,y,1-x-y)
 \nonumber \\
& + 
& {{g}^2  \over \pi} \int_{0}^{x+y} dz {x+2y-z \over 2(x -z)^2\sqrt{y
(x+y-z)}} \{  f^{3}_{\downarrow\uparrow\downarrow}(x,y,1-x-y)  \nonumber \\
&&  - f^{3}_{\downarrow\uparrow\downarrow}(z,x+y-z,1-x-y) \}
\nonumber \\ &
+ &{{g}^2  \over \pi x }\left( \sqrt{1 + {x\over y} } -1 \right) 
f^{3}_{\downarrow\uparrow\downarrow}(x,y,1-x-y)
 \nonumber \\
& + &{{g}^2  \over \pi}
\int_{0}^{1-x} dz
{f^{3}_{\downarrow\uparrow\downarrow}(x,z,1-x-z)\over (1-x) \sqrt{y z} } 
\nonumber
\\
& + &
{{g}^2  \over \pi} \int_{0}^{x+y} dz
{f^{3}_{\downarrow\uparrow\downarrow}(x+y -z ,z,1-x-y)  \over (x+y) 
\sqrt{y z} } \nonumber \\
 &  + &{h \over \sqrt{y}} f^{2}_{\uparrow\downarrow}(x+y,1-x-y) 
\delta(x) 
\nonumber \\ 
& - &{h \over \sqrt{y}} f^{2}_{\downarrow\uparrow}(x,1-x) \delta(1-x-y) 
\label{three}
\end{eqnarray}
and the  same equations with 
$\uparrow \leftrightarrow \downarrow$, $h \leftrightarrow -h$.
Note that $M^2$ is the eigenvalue of $2P^-$ is the normalisation of this 
subsection.
$m_b$ is the gluon mass $m_0$ after renormalisation 
resulting from normal ordering of gluon fields in $P^-$.
Although not necessary for our purposes of demonstrating pseudoscalar-vector
 splitting, 
for generality
we added a quark `kinetic' mass $m_f$ in the $f^2$ sector. 
Such counterterms in the 
hamiltonian should be allowed since
DLCQ and a one-gluon truncation break symmetries
such as parity \cite{burk}. 
Most importantly for us, a quark kinetic mass term does not 
break quark helicity symmetry.

\subsection{DLCQ gauge theory solution}
\label{sub:solve}
The endpoint delta functions of momentum, seen in section~\ref{sub:I2},
now  become spread out by the additional
interactions in the gauge theory. 
The $I$ interaction, that acts at endpoints only, 
couples to a part of the wavefunction of measure zero. 
One therefore expects the 
direct effects of $I$ in splitting the $0^{-+}$ and $0^{--}$ to vanish
as DLCQ $K \to \infty$. However, $I$ can combine with other 
interactions
to produce helicity-flip 
effects away from the endpoints. 

\begin{figure}
\centering
\BoxedEPSF{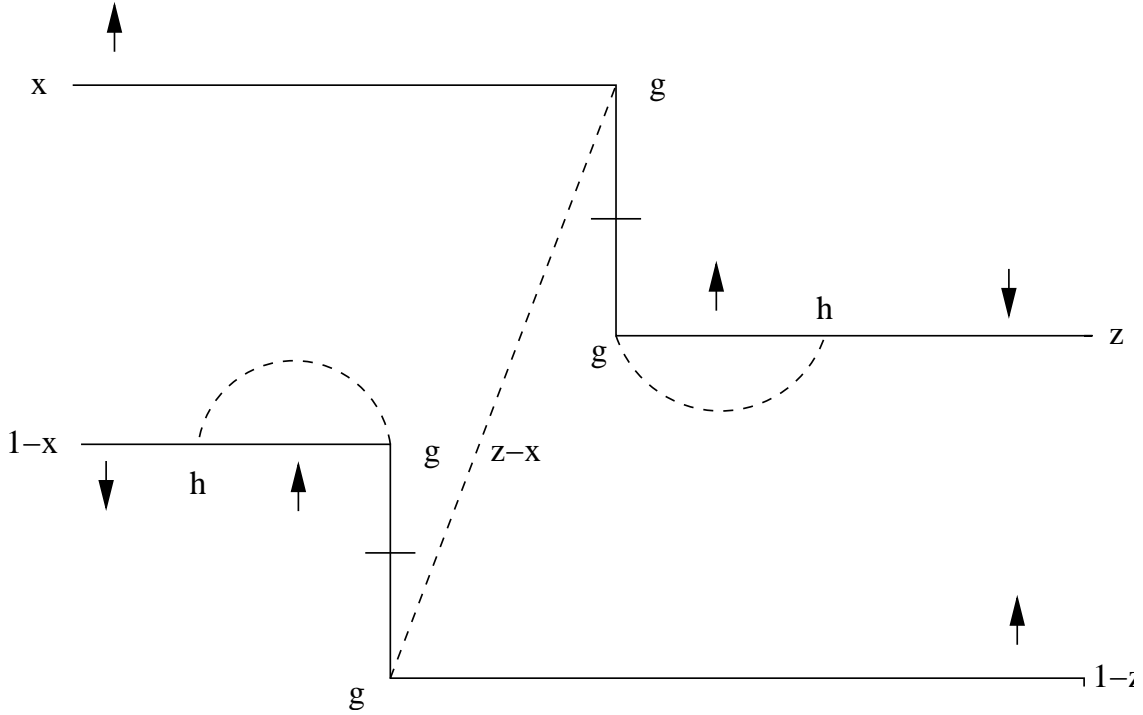 scaled 700}
\caption{Quark helicity flip process involving $I$. Solid lines are
quarks ($x^+$-instantaneous when barred), chain lines are dimensionally
reduced gluons. Vertices are
labelled by their corresponding couplings.}
\end{figure}

An example at
order $h^2 {g}^4$ is illustrated in Figure~3. 
The value that this process contributes to the expectation value of $M^2$
can be calculated in light-cone perturbation theory:
\begin{equation}
h^2 \left({{g}^2 \over \pi}\right)^2 \int_{0}^{1} dx \int_{x}^{1} dz  
{f^{2}_{\uparrow\downarrow}(x,1-x) f^{2}_{\downarrow\uparrow}(z,1-z)
 \over (1-x)^2 \left(M^2 - {m_{b}^{2} \over 1- x}\right) 
  (z-x) \left(M^2 - {m_{b}^{2}\over z-x}\right)
 \left(M^2 - {m_{b}^{2} \over z}\right)z^2} 
\end{equation} 
If $f^{2}$ and $M^2$ are finite, then this 
contribution is finite for finite $h$.

We solved numerically in DLCQ the dimensionally reduced QCD boundstate 
equations truncated to at most one gluon 
for the range $K=5$ to $K=30$. The particular
choice of the parameters $m_b, m_f, {g}, h$ is not very important,
since we did not try to tune them to obtain the best phenomenology.
However, they were zoned to ensure absence of tachyons, that the $0^{-\pm}$
states were  light in the spectrum, and a reasonably large
splitting of the $0^{-\pm}$ states compared to their masses.
A typical result for the spectrum in this case
is shown in Figure~4, where $M^2$ eigenvalues have been fit to the form
\begin{equation}
M^2 = A + {B \over \sqrt{K}} + {C \over K} + {D \over K^2} \ .
\end{equation}
The graph  shows that the splitting of the $0^{- \pm}$ survives
the DLCQ continuum limit $K \to \infty$, as advertised.
 
\begin{figure}
\centering
\BoxedEPSF{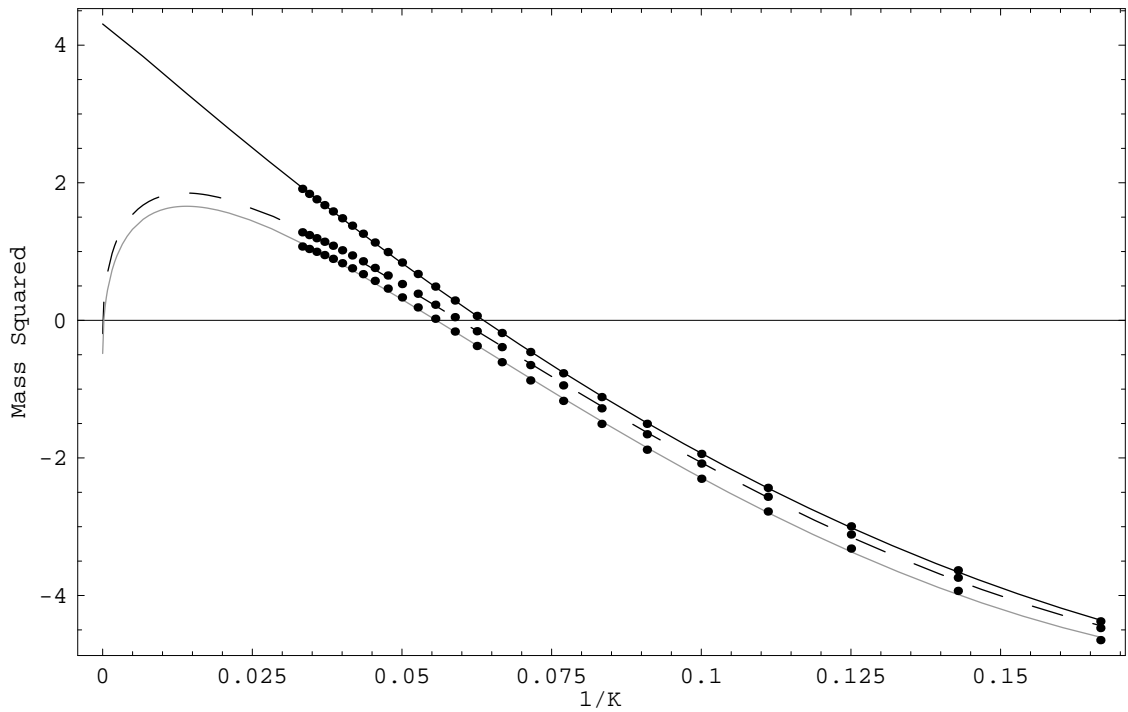 scaled 900}
\caption{Variation of mass squared $M^2$ with DLCQ cutoff $K$ for the lightest
three energy levels. The raw DLCQ data is smoothly extrapolated to $K=\infty$:
dark solid line $0^{--}$; chain line $0^{-+}$. For completeness we show
also the $1^{--}$ (gray line), which appears anomalously light for this 
particular parameter choice.}
\end{figure}

\section{Conclusions}
\label{sec:last}

We have shown that operator-valued integration constants, 
which arise in the solutions of the constraint equations encountered 
 in quantizing QCD in the light-cone representation, can 
dress the bare light-cone
 vacuum and induce new interactions.  
Without specifying the vacuum explicitly, 
except for the general form following from
 Poincare and gauge symmetry 
and the assumption that the bare vacuum {\em is} dressed by a pure fermion
component, we have derived 
the form of  new operators induced into the lightcone QCD hamiltonian.  
Those operators violate
quark helicity for zero quark bare mass.  
We have demonstrated qualitatively, and quantitatively
in the case of  dimensionally reduced
models, how that operator leads to a splitting of the masses of the
pseudoscalar and vector mesons, 
the helicity zero components of which would be degenerate otherwise.
The induced operators we have derived in this paper are almost certainly not
the only ones in QCD.  We have not analysed the gluon structure of the
vacuum  nor the gluon zero mode contribution to the hamiltonian, which is a 
separate study of interest. 

The illustrative DLCQ calculations we carried out for reduced models
can in principle be extended to $3+1$ dimensions via, for example,
transverse lattices \cite{tlreview,tlcalc} or the Pauli-Villars regulated 
formulation \cite{pfp}, where similar
induced operators will arise to spontaneously break quark helicity symmetry.
These topics are
now under investigation. Finally we draw the readers' attention to other 
recent work on chiral symmetry in light-cone field theory using path integrals
\cite{lenz} and importing chiral symmetry breaking from quantization 
on a space-like 
surface by rotating to the characteristic after having 
analyzed the chiral symmetry 
breaking \cite{pfp05}.

\section*{Acknowledgments}
Acknowledgments: The work of SD is supported by PPARC grant 
PPA/G/0/2002/00470.  The work of GM is supported by the U. S. 
Department of Energy through
contract DE-FG03-95ER40908. We thank B. van de Sande for helpful
discussions.

\appendix
\section{The Schwinger Model}
\label{sec:SM}

One example for which the vacuum has been derived
exactly is the Schwinger model (massless QED in
$1+1$ dimensions).  There exists a complete operator solution
which can be evaluated either at $x^0 = 0$ or at $x^+ = 0$~\cite{nm00}. 
Since there is one operator solution, if it has a condensate when quantized
at equal time there must also be the same one 
when quantized on the light cone. 
Here, we shall simply review those results, 
referring to the 
literature~\cite{nm00,mc94,mc99} for details and derivations.  
The action is 
\begin{equation}
S = \int d^2 x  \left[ -{1\over 4} \Tr \ F_{\alpha \beta} F^{\alpha\beta}
+ {\rm i}\bar{\psi} \gamma^{\alpha}_{(2)} D_{\alpha} \psi \right] \ ,
\end{equation}
$\alpha, \beta \in \{ + , -\} $,
$F_{\alpha \beta}$ is the field strength, $ \gamma^{\alpha}_{(2)}$
the two-dimensional representation of the gamma matrices, 
and $D_{\alpha}=\partial_{\alpha} + {\rm i}eA_{\alpha}$.
In light-cone gauge ($ A_{-} =0$), $\psi_-$ is the left moving part of a 
free massless fermi field.  
\begin{equation}
   \psi_- = \psi_-^0(x^+) .\label{pmeq}
\end{equation}
That is true however the theory is quantized.  If it is 
quantized on $x^+ = 0$, it is the only solution to the constraint 
equation 
\[
  \partial_- \psi_- = 0,
\]
The bosonized form is
\begin{equation}
         \psi_-^0 =  Z_-{\rm e}^{\Lambda_-^{(-)}}\sigma_-
 {\rm e}^{\Lambda_-^{(+)}} ,
\end{equation}
where $\Lambda_-$ is a bosonic field depending on $x^+$ (it is 
composed entirely of unphysical operators and physical states do not 
contain quanta from it), $\sigma_-$ is a 
(space-time independent) spurion and $Z_-$ is a wave function renormalization
 constant. (+) and (-) refer to positive and negative frequency
parts.  
We write $\psi_+$ similarly in bosonized form as 
\begin{equation}
     \psi_+ = Z_+ {\rm e}^{\Lambda_+^{(-)}}\sigma_+ {\rm e}^{\Lambda_+^{(+)}} 
\ .
\label{ppc}
\end{equation}

The physical vacuum in this theory is known exactly:
\begin{equation}
             \ket{\Omega(\theta)}  \equiv  \sum_{M=-\infty}^\infty 
{\rm e}^{{\rm i}M\theta}\ket{\Omega(M)} \quad;\quad  \ket{\Omega(M)} = 
({\sigma}_+^*{\sigma}_-)^M \ket{0} , \label{vac}
\end{equation}
where $\sigma^{-1} \equiv \sigma^*$.  This same state is found 
whether the system 
is quantized on $x^+ = 0$ or on $x^0 = 0$.  
The existence of these vacua, and their form, is determined from the fact
 that we have residual gauge transformations in lightcone gauge
and that ${\sigma}_+^*{\sigma}_-$
 is their generator. 
The vacuum  has the chiral condensate 
\begin{equation}
     \bra{\Omega(\theta)} {\bar{\psi} \psi} \ket{\Omega(\theta)} = -
 \frac{e}{2 \pi^{3/2}} e^\gamma \cos{\theta} .
\end{equation}
If the bare mass is nonzero there exists an induced interaction \cite{mc99}.

The case of the Schwinger model quantized on the light-cone with 
antiperiodic boundary conditions in $ x^-$ has also received a 
thorough discussion in the literature \cite{mc94,mc00}. 
The vacuum is exactly known also for 
Adjoint $QCD_{1+1}$, where again it is determined by residual gauge invariance
\cite{mrp}.

\end{document}